\documentclass[11pt]{article}
\usepackage[margin=1in]{geometry}
\usepackage{booktabs}
\usepackage{graphicx}
\usepackage{amsmath}
\usepackage{url}
\usepackage[hidelinks]{hyperref}
\usepackage{caption}
\date{September 12, 2026}

\title{Where Post-Training Quantization Breaks Text Embedders:\\
A Measured Map Across Four Embedder Families}
\author{Hyojung Han\\ThakiCloud}

\usepackage{placeins}
\begin{document}
\maketitle

\begin{abstract}
Weight-only post-training quantization (PTQ) is the cheapest way to shrink a retrieval
embedder, and the received advice for applying it --- protect the embedding table, allocate
bits by module sensitivity, and prefer a ranking-aware objective over weight reconstruction
--- comes from a decade of network and Transformer quantization work and has been carried
into LLM PTQ largely intact. We test that advice on retrieval embedders directly. We quantize five checkpoints drawn from four
architecture families across a grid of bit widths and group sizes, with symmetric
quantization as a control on a subset, and evaluate every arm on three retrieval corpora with per-query scores retained.
Two of the five checkpoints share a lineage on purpose: Qwen3-Embedding-0.6B and a
task-finetuned derivative of that exact checkpoint form a paired control, and we do not
count them as independent evidence of generalization. Generalization rests on the four
families.

Three operating regimes emerge, and together they challenge key parts of that advice. At INT4/g16
there is almost nothing for a mixed-precision allocator to redistribute: quantizing the
embedding table, the attention projections, or the feed-forward block in isolation costs at
most 1.01 NDCG points on any individual model--corpus pair, and the sign of that cost is not
consistent. We report per-query paired bootstrap intervals rather than claiming these effects
are statistically zero --- on our largest corpus, 4{,}392 queries resolve differences of a few
tenths of a point, and ten of forty-five module cells do exclude zero. The argument is about
magnitude, not significance: the module-level headroom we observe at this operating point is
about one NDCG point. At INT3 module sensitivity becomes large enough to matter, but its
ordering is family-dependent and the costs stop being additive --- the feed-forward block is
costliest in the Qwen lineage while attention dominates BGE-M3 and E5, EmbeddingGemma switches
between the two with the group size, and the interaction residual varies in both magnitude and sign. At INT2/g16 checkpoints whose
reconstruction error is clustered between 0.315 and 0.337 retain between 1.3\% and 65.9\% of
their retrieval quality.

Cutting across the three regimes,
weighted weight-reconstruction error is a good predictor of retrieval loss along one axis and
a poor one along another: it tracks damage from \emph{uniform} quantization closely
($r=0.80$ to $0.92$ in each of the five checkpoints) but tracks damage from quantizing
\emph{individual modules} weakly: computed within an operating point, where the allocator's
question actually lives, it runs from $-0.01$ at INT4 to $0.41$ at INT3, and is $0.42$ at the
INT2 cliff. The weak axis is the one a mixed-precision
allocator searches. And decisively, that same predictor fails \emph{across} model families exactly
where the decision matters: at reconstruction errors clustered between 0.315 and 0.337,
INT2 leaves one model with 66\% of its retrieval quality and another with 1\%. The low-bit cliff is strongly model-family
dependent, and no positive protection priority between attention and the
feed-forward block transfers reliably across all four families. The one cross-family
regularity is negative, and it survives the cliff: the embedding table never emerges as the dominant
isolated protection priority, even at INT2. We do not claim
architecture causes it: the families differ in parameter count, pretraining objective and
training data as well as in architecture. We show that vocabulary size and embedding-table fragility cannot by
themselves explain the cliff: quantizing only the embedding table to INT2 retains 97.6\% to
100.5\% of retrieval quality in the per-model averages (94.7\% to 101.3\% across individual
model--corpus cells), including in the three checkpoints whose full-model INT2 retention is
below 2\%. Tokenization behaviour itself is untested. The paired control rules out the obvious alternative explanation: task fine-tuning
on 127{,}190 pairs moves quantization tolerance by at most 1.0 point in two-corpus averages
(2.28 points on any individual arm), and the INT3 decline
and INT2 collapse are both already present in the base checkpoint.

All evaluated checkpoints are publicly available; the ones we produced --- SKILLRET-Embedding-0.6B
and the distilled SKILLRET-Edge-22M and 109M with their quantized variants --- are collected
at \url{https://huggingface.co/ThakiCloud}, and the other four are the upstream releases cited
in Section~\ref{sec:protocol}. The full measurement repository (\url{https://github.com/ThakiCloud/skillret-ptq-measurements})
will be made public with the arXiv version, carrying the arm-by-arm measurements with per-query scores,
the byte-verified size ledger, and a prompt-contract gate derived from a measured failure in
which a train/eval prefix mismatch manufactured a spurious result that quantization
outperforms full precision.
\end{abstract}

\section{Introduction}

A retrieval embedder is a natural target for compression. It is small relative to a
generative model, it is called once per query and once per document, and in agent systems it
sits on the critical path of every turn. Weight-only PTQ is attractive because it needs no
training data and no gradient step: quantize the checkpoint, serve it, done.

The practical questions are which bit width survives, and whether spending extra bits on
selected modules buys anything. Both questions have well-known answers for generative
LLMs. GPTQ~\cite{gptq} compensates quantization error layer by layer against a calibration
Hessian; AWQ~\cite{awq} identifies and protects salient channels;
OmniQuant~\cite{omniquant} optimizes clipping and scaling; mixed-precision
allocators~\cite{impq} distribute bits by an estimated per-layer cost. The shared premise is
that some parts of the network matter more, and that a cheap proxy for how much they matter
can be measured and then allocated against.

Whether that premise holds for a \emph{retrieval} embedder is a separate question, because
the success criterion is different. A generative model is judged by perplexity, an average
over token distributions. An embedder is judged by whether the top-$k$ documents for a query
come back in the right order, which depends only on the relative geometry of a handful of
scores near a decision boundary. Averaged reconstruction error and top-$k$ order are not
obviously the same quantity, and the gap between them is the stated motivation for
retrieval-aware compression objectives.

We set out to build such an objective. Before doing so we measured the premise, and the
measurement did not support it. This paper reports what we found instead.

The measurements separate into three regimes, and we organise the paper around them. At
INT4/g16 quantization is NDCG-neutral and no module carries a sensitivity differential worth
allocating against; widening the group to 128 already costs more. At INT3 sensitivity appears, module costs stop being additive, and the
ordering of which module matters is family-specific. At INT2 checkpoints at nearly identical
reconstruction error separate by tens of points of retained retrieval quality.

\paragraph{Contributions.}
\begin{enumerate}
\item \textbf{Three regimes, measured across four embedder families.} Five checkpoints, three
corpora, a grid over bit width and group size, with per-query scores retained. At INT4/g16 quantization
is NDCG-neutral and no module carries a sensitivity differential large enough to allocate against: the
largest single-module effect on any individual model--corpus pair is 1.01 points. At INT3 sensitivity appears, but its
ordering inverts between families --- at INT3/g16 the feed-forward block is the costliest
module for Qwen3 and its fine-tuned derivative, attention is costliest for EmbeddingGemma,
BGE-M3 and E5, and EmbeddingGemma switches to the feed-forward block at g32 --- and module costs stop
being additive, with the
interaction residual varying in magnitude and, at INT3/g16, in sign. At INT2, checkpoints whose
reconstruction error is clustered between 0.315 and 0.337 retain between 1.3\% and 65.9\% of
their retrieval quality. At the cliff itself, embedding-only quantization still retains
97.6\% to 100.5\% across all five checkpoints, while the dominant isolated damage shifts to
the feed-forward block in the Qwen lineage and to attention in BGE-M3.

\item \textbf{The standard sensitivity proxy holds on one axis and fails on the other.}
Weight reconstruction error tracks the damage from uniform quantization well ($r=0.80$ to
$0.92$ in each of the five checkpoints, Spearman $\rho\geq0.89$ in all five and $1.00$ in two) and tracks the damage from
quantizing individual modules poorly ($r=-0.04$ to $0.42$ across fixed operating points; pooling
module arms across bit widths inflates this to $0.66$ by range extension). The weak axis is
the one a mixed-precision allocator searches. Module-sensitivity policies are therefore not
reliably transferable across embedder families, which is a constraint on what any such method
can promise rather than a demonstration that a particular allocator fails.

\item \textbf{NDCG-neutral is not ranking-invariant, measured against the relevance
judgements.} At INT4/g16, where full-model NDCG changes by less than one point in every
per-model three-corpus average, 13\% to 22\% of the top-10 is
replaced and the boundary document moves for about half of queries. Counted against qrels
rather than inferred from the metric, 97.8\% of those evictions are non-relevant pooling all
evictions and 94.8\% weighting corpora equally, relevant-only overlap stays at 0.944--0.970
in the per-model averages (0.899--1.000 across individual model--corpus cells) and
Recall@10 moves by under a point in the per-model averages, and by at most 2.17 points on
any individual model--corpus cell. The protection is bounded: before a model's own cliff the churn stays
in the non-relevant tail and the gold set survives, and past it the gold set is what fails ---
relevant-only overlap falls to 0.001--0.006 for the collapsing checkpoints while the
non-relevant share of evictions stays high. That is what distinguishes surviving INT2 from
collapsing at it.

\item \textbf{Controls, and a reproducibility contract built from measured failures.} A paired
base/fine-tuned control shows task adaptation changes quantization tolerance by at most 1.0
point in two-corpus averages, and 2.28 points on any individual arm, and creates neither the INT3 decline nor the INT2 collapse. An intervention on pooling
refutes the sufficiency of the most attractive explanation for the cliff and leaves its
necessity undecided, so we report the cliff as unexplained. The prompt-contract, byte-ledger and
holdout-isolation gates each exist because the corresponding failure produced a number we
briefly believed.
\end{enumerate}

The distilled-student comparison in Section~\ref{sec:discussion} is a practical consequence of
these measurements rather than a contribution of its own.

\section{Related Work}

\subsection{Mixed-precision and post-training quantization}

The idea this paper tests --- measure how sensitive each part of a network is, then spend
bits accordingly --- is older than LLM quantization. HAQ~\cite{haq} searched hardware-aware
mixed-precision assignments; HAWQ~\cite{hawq} used the Hessian spectrum as a layer
sensitivity signal and HAWQ-V2~\cite{hawqv2} replaced the top eigenvalue with the Hessian
trace and selected precision along a Pareto frontier. On the reconstruction side,
AdaRound~\cite{adaround} showed that rounding is a choice rather than a given,
BRECQ~\cite{brecq} reconstructs at block granularity precisely because layers are not
independent, and QDrop~\cite{qdrop} attacks the flatness of the loss surface at extremely low
bit widths. LAPQ~\cite{lapq} is the closest precedent for one of our observations: it reports
that the quantization loss is close to separable under mild quantization and becomes
non-separable as quantization grows aggressive.

The LLM-era methods inherit this line. GPTQ~\cite{gptq} compensates error layer by layer
against a calibration Hessian, AWQ~\cite{awq} protects salient channels,
SmoothQuant~\cite{smoothquant} and ZeroQuant~\cite{zeroquant} move difficulty between weights
and activations, OmniQuant~\cite{omniquant} optimizes clipping and scaling, and
CoopQ~\cite{impq}, which circulated as IMPQ, allocates mixed precision with a cost that
models interaction between layers rather than treating them independently.

We therefore do not claim that quantization interactions, or their effect on bit allocation,
are a new observation. Our question is narrower and, as far as we can tell, untested: whether
the sensitivity signals this literature relies on remain usable when the quantity being
preserved is \emph{retrieval order} rather than reconstruction error or perplexity, and
whether the resulting policy is the same for embedders of different families.

\subsection{Low-bit Transformer encoders}

Encoder Transformers have their own low-bit lineage, and it is the closer one: three of our
five checkpoints are encoders. Q8BERT~\cite{q8bert} established 8-bit BERT;
Q-BERT~\cite{qbert} combined Hessian-based mixed precision with group-wise quantization and
pushed to two bits; TernaryBERT~\cite{ternarybert} and BinaryBERT~\cite{binarybert} reach
ternary and binary weights; I-BERT~\cite{ibert} removes floating point from inference
entirely. Q-BERT in particular anticipates both the method and the granularity we use.

The distinction that matters here is the training budget. Every result above involves
fine-tuning, quantization-aware training, or distillation. We quantize existing checkpoints
and train nothing, which is the regime a practitioner is in when a good embedder already
exists and only has to be made smaller.

\subsection{Extreme low-bit LLMs}

Two- and three-bit weight quantization is well studied for generative LLMs, with useful
quality demonstrated by specialized methods, which makes our INT2 results look
anomalous until the methods are compared. SpQR~\cite{spqr} and SqueezeLLM~\cite{squeezellm}
keep a sparse high-precision residual; QuIP\#~\cite{quipsharp} and AQLM~\cite{aqlm} replace
scalar quantization with lattice and additive codebooks; QuaRot~\cite{quarot} and
SpinQuant~\cite{spinquant} rotate away outliers before quantizing; BitNet~\cite{bitnet158}
trains in 1.58 bits from the start. None of these is uniform group-wise scalar PTQ applied to
a finished model, which is what we measure. Their success at two bits is therefore not
evidence that our arms should have survived, and our collapse is not evidence against them.

\subsection{Low-bit text embedders}

Two efforts target embedder weights directly. EmbeddingGemma~\cite{embgemma} ships
quantization-aware-trained INT4 and INT8 mixed-precision checkpoints. BitNet Text
Embeddings~\cite{bitembed} converts Qwen3 and Gemma3 backbones to ternary weights and
recovers quality through continual contrastive pre-training and distillation from a
full-precision teacher. Both change the checkpoint. Our one surviving ternary control does not contradict
them: ternary embedders work when they are trained to be ternary, which is a different claim
from an existing embedder surviving ternary PTQ.

\subsection{Compressing representations rather than weights}

A larger literature compresses what the encoder outputs. Product quantization~\cite{pq} and its
optimized variants~\cite{opq} underpin billion-scale search, anisotropic vector
quantization~\cite{avq} reshapes the quantization error to favour inner-product recall, and
binary passage retrieval~\cite{bpr} hashes passages outright. JPQ~\cite{jpq} and
RepCONC~\cite{repconc} are the closest in spirit to a retrieval-aware objective: both train
the query encoder jointly with the index under a ranking loss rather than a reconstruction
loss. Recent work continues in this direction, including learning-free binary embeddings
built from an isolation kernel~\cite{ike}, alongside analyses of how far output-vector
precision can be reduced before top-$k$ retrieval degrades~\cite{point1,topktheory} and
contextual quantization for re-ranking~\cite{ctxquant}.

This line shares our metric and not our object. It compresses the index, under a storage
budget for vectors; we compress the model, and hold the index at full precision throughout so
that every number here is attributable to model weights. JPQ and RepCONC also matter for a
second reason: they are existing evidence that a ranking-aware objective helps when the thing
being quantized is the representation. Whether that carries over to the weights is the
question our measurements bear on.

\section{Experimental Protocol}
\label{sec:protocol}

\paragraph{Models.} Four architecture families, five checkpoints.
Qwen3-Embedding-0.6B~\cite{qwen3emb} is a causal decoder with last-token pooling;
EmbeddingGemma-300M~\cite{embgemma} is a bidirectional encoder with mean pooling; BGE-M3
\cite{bgem3} is an XLM-R-family multilingual encoder; and E5-base-v2~\cite{e5} is a
BERT-family encoder, the smallest model here at 110M parameters. The fifth checkpoint,
SkillRet-Embedding-0.6B, is \emph{not} a fifth family: it is Qwen3-Embedding-0.6B fine-tuned
on 127{,}190 query--skill pairs with a multiple-negatives ranking objective. We include it as
a \emph{paired control} that varies training while holding architecture, tokenizer, and
hidden size fixed, and we exclude it when counting evidence for cross-family claims.

\paragraph{Corpora.} SciFact and NFCorpus from BEIR~\cite{beir}, and the public SkillRet
test split~\cite{skillret}, an agent skill-retrieval task whose queries carry one to three
gold documents. We report NDCG@10~\cite{ndcg}, the headline metric of all three.

We pin the SkillRet split explicitly, and we pin the version of the benchmark paper we pin it
against. We evaluate on the \emph{published} dataset \texttt{ThakiCloud/SkillRet} on the
Hugging Face Hub, whose test split contains \textbf{4{,}392 queries over 6{,}006 documents};
we verified our local copy is byte-identical to the published files at revision
\texttt{a050ad2}. The Hub is mutable, so the revision is what the numbers here are pinned to. The benchmark
paper~\cite{skillret} has been revised: v1 described 17{,}810 skills and 4{,}997 evaluation
queries, while v2 onward describes 16{,}129 and 4{,}392, and v3 additionally reordered the
author list. Our query count matches v2 and later. Figures quoted against the v1 split are
therefore not comparable to ours, and we do not convert between the two.

\paragraph{Quantization.} Group-wise affine (asymmetric) min--max quantization in the sense
of~\cite{jacob}, with granularity in the sense of~\cite{krish}; scales and zero-points are
stored in FP16, matching a GGUF-style block. The symmetric arm drops the zero-point and uses
absmax. Weights only: activations remain FP32. These experiments characterize weight-quantization
quality; they do not establish integer-kernel latency or end-to-end deployed accuracy, both
of which require runtime-specific validation. A weight-only packed kernel need not quantize
activations at all, so we make no claim about the sign of the gap. We never report a
fake-quantized score alongside a kernel latency.

\paragraph{Prompt contract.} Each model carries its own query prompt convention. We load
every model through its own \texttt{sentence-transformers} configuration so that pooling and
prompt are the model's own, we record the resolved prompt in the result file, and we use one
identical convention across all arms of a model. Section~\ref{sec:repro} explains why this is
a gate and not a detail.

\begin{table}[t]\centering\small
\caption{Full-precision (FP32 weights) NDCG@10 for every model and corpus. These are the
baselines every retention figure in this paper is relative to; absolute values differ across
models partly because SkillRet is in-domain for the finetuned checkpoint only.}
\label{tab:protocol}
\begin{tabular}{lrrr}
\toprule
Model & NFCorpus & SciFact & SkillRet \\
\midrule
Qwen3-Emb-0.6B & 35.40 & 70.32 & 57.48 \\
SkillRet-0.6B (finetuned) & 33.86 & 69.43 & 78.48 \\
EmbeddingGemma-300M & 39.04 & 78.61 & 61.80 \\
BGE-M3 & 31.57 & 64.37 & 56.25 \\
E5-base-v2 & 35.59 & 71.94 & 52.95 \\
\bottomrule
\end{tabular}\end{table}

\section{Results}

\begin{figure}[t]\centering
\includegraphics[width=\linewidth]{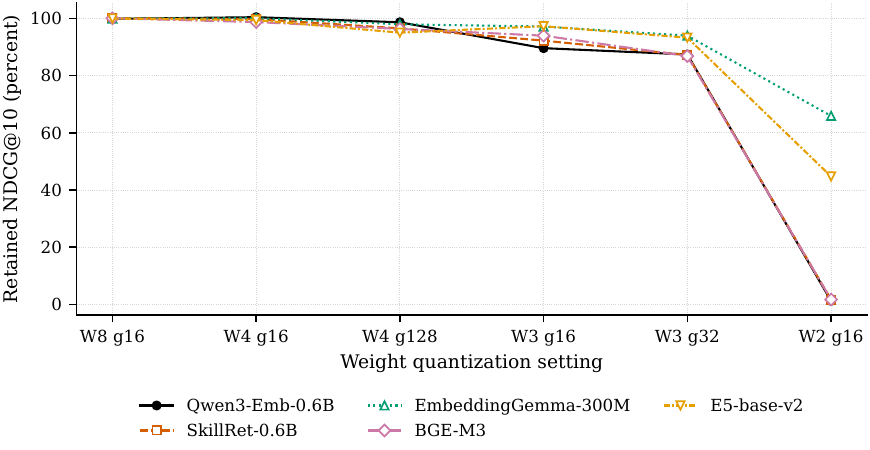}
\caption{Retained NDCG@10 as a percentage of the full-precision score, averaged over
SciFact, NFCorpus and SkillRet, under uniform group-wise affine weight quantization. All
points use group size 16 except the W4/g128 control. W8 and W4/g16 are approximately
NDCG-neutral for every model; widening the group to 128 costs more in some checkpoints. The
models separate at W3 and diverge at W2, where EmbeddingGemma retains roughly two thirds of
its retrieval quality and the Qwen-lineage models retain almost none. Ternary is not plotted
here --- only one of its five arms survived the symmetric-quantizer retraction
(Table~\ref{tab:uniform}).}
\label{fig:ladder}\end{figure}

\begin{table}[t]\centering\small
\setlength{\tabcolsep}{4pt}
\caption{Retained NDCG@10 as a percentage of the full-precision score, averaged over the
three corpora. Weight-only quantization; group-wise affine unless marked \emph{sym}.
Bold marks arms that lost more than half their retrieval quality. W8/g16 and W4/g16 are
approximately NDCG-neutral for every model; widening to g128 already costs E5 4.4 points of
retention and the fine-tuned checkpoint 3.3 --- which is not the same as harmless; see
Table~\ref{tab:turnover}. The models separate at W3 and diverge at W2: EmbeddingGemma retains roughly
two thirds of its quality where the two decoder-lineage and XLM-R-lineage models retain
almost none. Ternary is reported for E5 only: the ternary and symmetric arms for the other
four checkpoints were measured but retracted after a defect was found in the symmetric
quantizer (Section~\ref{sec:repro}), and were not re-measured. One surviving ternary arm
is not evidence about the other four.}
\label{tab:uniform}
\resizebox{\linewidth}{!}{%
\begin{tabular}{lrrrrrrrr}
\toprule
Model & W8 g16 & W4 g16 asym & W4 g16 sym & W4 g128 asym & W3 g16 & W3 g32 & W2 g16 & Ternary \\
\midrule
Qwen3-Emb-0.6B & 99.8 & 100.4 & -- & 98.7 & 89.6 & 87.4 & \textbf{1.3} & -- \\
SkillRet-0.6B (finetuned) & 100.1 & 99.8 & -- & 96.5 & 92.2 & 87.2 & \textbf{1.7} & -- \\
EmbeddingGemma-300M & 100.0 & 100.1 & 98.9 & 98.0 & 97.2 & 94.0 & 65.9 & \textbf{1.4} \\
BGE-M3 & 100.1 & 98.7 & -- & 96.3 & 94.0 & 86.8 & \textbf{1.8} & -- \\
E5-base-v2 & 99.9 & 99.4 & 99.5 & 95.0 & 97.3 & 93.3 & \textbf{44.8} & \textbf{1.9} \\
\bottomrule
\end{tabular}}\end{table}

\subsection{Module sensitivity appears late, and its ordering does not transfer reliably}
\label{sec:modules}

\begin{table}[t]\centering\small
\caption{Change in NDCG@10 (points, relative to full precision, averaged over three
corpora) when \emph{only} the named module is quantized, against quantizing everything.
Three things vary. At INT4/g16 isolated-module effects stay small --- at most 0.59 in the
averages shown here, and 1.01 on any individual model--corpus pair --- leaving little
sensitivity differential to allocate against. At INT3 sensitivity appears, but the ordering
is not shared: at g16 the feed-forward block is the most costly module for Qwen3 and its
fine-tuned derivative, BGE-M3 reverses the ordering with a free FFN and attention as the
bottleneck, E5 is attention-dominant at both group sizes, and EmbeddingGemma's costliest
module is attention at g16 and the feed-forward block at g32. And in every model
the parts do not add up to the whole. The last column is the interaction residual
$I=\Delta_{\text{joint}}-\sum\Delta_{\text{module}}$, shown only where all three
single-module arms were measured; negative means joint quantization hurts more than the parts
predict. At INT4/g16 $I$ is small ($-0.15$ to $+0.17$); at INT3 it is substantial in every
model and at g16 it changes sign between them, so a bit allocator that
treats per-module sensitivity as an independent cost is wrong by a model-dependent amount in
a model-dependent direction. The
embedding table is nearly free everywhere, at every operating point.}
\label{tab:modules}
\begin{tabular}{lrrrrr}
\toprule
Model & Embed.\ only & Attn.\ only & FFN only & All & $I$ \\
\midrule
\multicolumn{6}{l}{\emph{INT4 / g16}} \\
\quad Qwen3-Emb-0.6B & +0.01 & +0.16 & +0.16 & +0.22 & -0.10 \\
\quad SkillRet-0.6B (finetuned) & +0.01 & -0.23 & +0.18 & -0.19 & -0.15 \\
\quad EmbeddingGemma-300M & -0.04 & +0.21 & +0.04 & +0.08 & -0.13 \\
\quad BGE-M3 & +0.06 & -0.59 & -0.10 & -0.67 & -0.04 \\
\quad E5-base-v2 & -0.23 & -0.25 & -0.01 & -0.32 & +0.17 \\
\multicolumn{6}{l}{\emph{INT3 / g16}} \\
\quad Qwen3-Emb-0.6B & -0.04 & -1.34 & -1.86 & -5.59 & -2.35 \\
\quad SkillRet-0.6B (finetuned) & -0.02 & -1.04 & -1.50 & -4.47 & -1.91 \\
\quad EmbeddingGemma-300M & -0.09 & -0.71 & -0.56 & -1.83 & -0.47 \\
\quad BGE-M3 & +0.16 & -2.46 & -0.01 & -3.11 & -0.80 \\
\quad E5-base-v2 & -0.32 & -0.98 & -0.36 & -1.28 & +0.39 \\
\multicolumn{6}{l}{\emph{INT3 / g32}} \\
\quad Qwen3-Emb-0.6B & -0.51 & -2.09 & -3.17 & -6.73 & -0.96 \\
\quad SkillRet-0.6B (finetuned) & -0.06 & -2.07 & -3.51 & -7.76 & -2.12 \\
\quad EmbeddingGemma-300M & -0.40 & -1.13 & -1.67 & -3.77 & -0.56 \\
\quad BGE-M3 & +0.27 & -5.09 & -1.14 & -6.64 & -0.68 \\
\quad E5-base-v2 & -0.34 & -1.81 & -0.04 & -3.50 & -1.31 \\
\bottomrule
\end{tabular}\end{table}
\begin{table}[t]\centering\small
\caption{Per-query paired bootstrap on the INT4 arms. For each model we show the corpus with
the largest absolute effect, its 95\% interval over 10{,}000 resamples of
$\Delta\text{NDCG}_q$, and how many of that model's three corpora have an interval
excluding zero. No single-module effect exceeds 0.59 in the three-corpus averages, or 1.01
on any individual model--corpus pair. Ten of the forty-five
module cells exclude zero, which is what 4{,}392 queries buys in resolution rather than
evidence of a usable sensitivity gap --- and four of those ten are positive.}
\label{tab:ci}
\begin{tabular}{lrcr}
\toprule
Arm & Worst $\Delta$ (pts) & 95\% CI & CI excludes 0 \\
\midrule
\multicolumn{4}{l}{\emph{Qwen3-Emb-0.6B}} \\
\quad Embed.\ only & +0.21 & $[+0.06, +0.35]$ & 1/3 \\
\quad Attn.\ only & +0.79 & $[+0.48, +1.12]$ & 1/3 \\
\quad FFN only & +0.18 & $[-0.74, +1.12]$ & 0/3 \\
\quad All & +0.88 & $[+0.46, +1.30]$ & 1/3 \\
\multicolumn{4}{l}{\emph{SkillRet-0.6B (finetuned)}} \\
\quad Embed.\ only & -0.12 & $[-0.70, +0.45]$ & 0/3 \\
\quad Attn.\ only & -0.31 & $[-1.10, +0.47]$ & 0/3 \\
\quad FFN only & +0.67 & $[-0.15, +1.52]$ & 1/3 \\
\quad All & -0.39 & $[-0.70, -0.09]$ & 1/3 \\
\multicolumn{4}{l}{\emph{EmbeddingGemma-300M}} \\
\quad Embed.\ only & -0.26 & $[-0.68, +0.06]$ & 0/3 \\
\quad Attn.\ only & +0.18 & $[-0.22, +0.61]$ & 0/3 \\
\quad FFN only & +0.50 & $[-0.18, +1.20]$ & 1/3 \\
\quad All & +0.23 & $[-0.54, +1.00]$ & 0/3 \\
\multicolumn{4}{l}{\emph{BGE-M3}} \\
\quad Embed.\ only & +0.15 & $[-0.06, +0.47]$ & 0/3 \\
\quad Attn.\ only & -1.01 & $[-2.08, +0.04]$ & 1/3 \\
\quad FFN only & -0.69 & $[-1.41, -0.04]$ & 2/3 \\
\quad All & -1.63 & $[-2.85, -0.44]$ & 1/3 \\
\multicolumn{4}{l}{\emph{E5-base-v2}} \\
\quad Embed.\ only & -0.28 & $[-0.93, +0.30]$ & 1/3 \\
\quad Attn.\ only & -0.48 & $[-1.19, +0.20]$ & 1/3 \\
\quad FFN only & -0.44 & $[-0.65, -0.23]$ & 1/3 \\
\quad All & -1.19 & $[-1.46, -0.91]$ & 1/3 \\
\bottomrule
\end{tabular}\end{table}

The standard recipe for mixed-precision PTQ is to measure each module's contribution to some
error, then distribute bits against those contributions. Table~\ref{tab:modules} shows three
reasons that recipe does not straightforwardly apply to these embedders.

First, at INT4 with group 16 there is almost nothing to distribute. Quantizing the embedding
table alone, the attention projections alone, or the feed-forward block alone costs at most
1.01 points in any model on any corpus. Quantizing all three jointly can cost somewhat more,
reaching 1.63 points, which is consistent with the non-additive interactions reported below. Per-query paired
bootstrap intervals (Table~\ref{tab:ci}) show that this is a statement about magnitude and
not about noise: thirty-five of forty-five module cells have a 95\% interval covering zero,
but the ten that do not are simply small effects measured on 4{,}392 queries, and four of
them are \emph{positive}. We do not use statistical significance to define sensitivity; as a
robustness check, two-sided bootstrap $p$-values computed from the empirical proportion of
resamples falling on either side of zero, then Benjamini--Hochberg corrected across the
forty-five cells at $q=0.05$
leaves eight, and the largest effect is unchanged at 1.01 points --- the same global maximum
quoted above, not a separate figure. Normalising each change by its
own full-precision score rather than reporting absolute points leaves every module ordering
in Table~\ref{tab:modules} intact in all five checkpoints. A bit allocator run at this operating point is choosing among near
equals, and it cannot even rely on the sign.

Second, sensitivity does appear at INT3 --- but the ordering is not shared. For
Qwen3-Embedding and its fine-tuned derivative the feed-forward block is the most expensive
module to quantize at both group sizes. For BGE-M3 it is the cheapest: at g16 the FFN costs
0.01 points while attention costs 2.46. The inversion is not an artifact of one setting --- at
INT3/g32 BGE-M3's attention costs 5.09 points against 1.14 for its FFN, the largest
single-module effect anywhere in our grid. E5-base-v2 is attention-dominant at both group
sizes (0.98 against 0.36 at g16, 1.81 against 0.04 at g32), and EmbeddingGemma does not keep
one ordering across group sizes: attention costs more at g16 (0.71 against 0.56) and the
feed-forward block at g32 (1.67 against 1.13). A policy that says ``protect the feed-forward
block'' is right for the Qwen lineage, right for EmbeddingGemma at one group size and wrong at
the other, and spends its extra bits on the wrong tensors on BGE-M3 and E5.

Third, and independently of which module wins, the per-module costs do not add up to the
joint cost. We report the interaction residual $I = \Delta_{\text{joint}} -
(\Delta_{\text{embed}} + \Delta_{\text{attn}} + \Delta_{\text{ffn}})$ rather than a ratio,
because a ratio explodes when the denominator approaches zero. At INT3/g16, $I$ is $-2.35$ for Qwen,
$-1.91$ for its fine-tune, $-0.80$ for BGE-M3 and $-0.47$ for EmbeddingGemma --- joint
quantization is worse than the parts predict --- but $+0.39$ for E5, where it is
\emph{better}. At INT3/g32 every model is negative, from $-0.56$ to $-2.12$. So the
interaction is substantial in all five models and at both operating points, its magnitude
varies four- to six-fold, and its sign is not even stable across models at one setting. Treating
per-module sensitivity as an independent cost, the assumption behind additive knapsack and
linear bit-allocation formulations, is therefore unsafe in a way that cannot be corrected by a
constant: an allocator calibrated on one model can be wrong about the direction of the error
on another.

The one regularity that does hold across every checkpoint is negative: the embedding
table is nearly free to quantize, costing at most 0.51 points at INT4 and INT3, despite
accounting for 21\% to 66\% of the parameters depending on the model.

We first stated that rule having measured module arms only at INT4 and INT3, which is to
say everywhere except the regime where models actually break. Section~\ref{sec:int2mod}
measures the cliff itself, and the rule survives in a stronger form: quantizing only the
embedding table to INT2 leaves 97.6\% to 100.5\% of retrieval quality in all five
checkpoints, including the three that keep under 2\% when everything is quantized. If there
is a transferable rule here, it is that the largest tensor is the one that never emerges as the
dominant isolated protection priority --- and that holds at the operating point where every
other rule we tested failed.

\subsection{Reconstruction error predicts one axis, not the other}
\label{sec:recon}

\begin{figure}[t]\centering
\includegraphics[width=\linewidth]{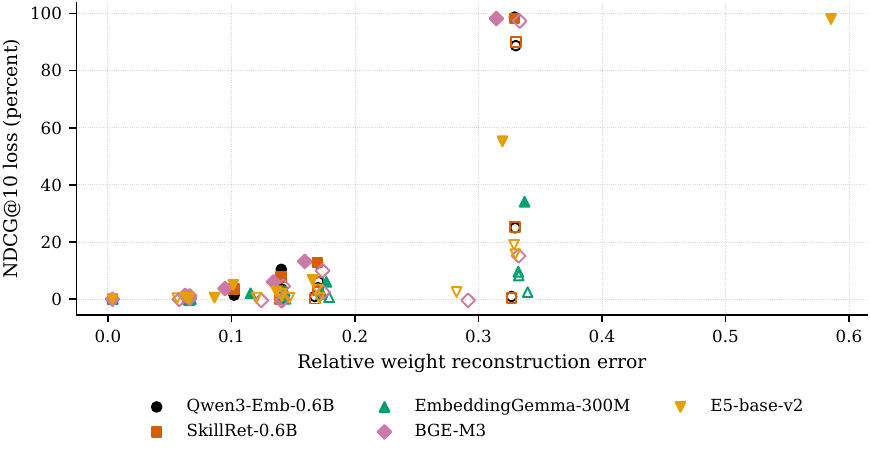}
\caption{NDCG@10 loss against weighted relative weight reconstruction error, one point per
quantization arm. Filled markers are uniform arms (bit width and group size swept over the
whole model); open markers of the same shape are single-module arms. The filled points rise
together, which is why reconstruction error is a usable predictor along that axis. The open
points do not: several sit at high reconstruction error and near-zero loss. Around
reconstruction error 0.30 to 0.34, arms from different families --- and isolated module arms
within them --- span almost the full range of retrieval loss.}
\label{fig:recon}\end{figure}

\begin{table}[t]\centering\small
\caption{Pearson correlation between parameter-weighted relative weight reconstruction
error (weighted over the quantized tensors) and NDCG@10 loss, excluding arms that lost more than half their quality. Left: computed
within each model and split by arm family. Reconstruction error tracks damage well along the
\emph{uniform} axis --- how hard the whole model was quantized --- and poorly along the
\emph{module} axis --- which part was quantized, the axis a mixed-precision allocator
actually searches. Middle: the module axis at fixed operating points, which
removes bit-width severity as a confound and asks whether reconstruction error tracks module
damage consistently across the evaluated checkpoints. Model-specific evidence for the same
point is the ordering inversion in Table~\ref{tab:modules}. Pooling module arms across bit widths
inflates the correlation to 0.658 by range extension, because severity then varies along with
module choice; within a bit width it never exceeds 0.42, and at INT4 it is approximately zero
($r=-0.043$). The INT2 module arms added in
Section~\ref{sec:int2mod} sit in the same weak band as INT3 rather than improving it.
Right: the INT2 arm per model, where nearly equal reconstruction error produces very
different retrieval outcomes. $^{\ddagger}$Pooled over all bit widths and inflated for the
reason above; the middle table is the figure to read. Every column drops arms losing more than
half their quality, which at INT2 removes three of fifteen module arms; relaxing that rule
moves the INT2 figure to 0.294 rather than improving it.}
\label{tab:recon}
\begin{tabular}{lrrr}
\toprule
Model & Uniform arms & Module arms$^{\ddagger}$ & All \\
\midrule
Qwen3-Emb-0.6B & 0.865 & 0.648 & 0.617 \\
SkillRet-0.6B (finetuned) & 0.92 & 0.622 & 0.595 \\
EmbeddingGemma-300M & 0.915 & 0.828 & 0.637 \\
BGE-M3 & 0.897 & 0.596 & 0.531 \\
E5-base-v2 & 0.804 & 0.828 & 0.791 \\
\midrule
Pooled & 0.875 & 0.658 & 0.631 \\
\bottomrule
\end{tabular}\quad
\begin{tabular}{lrr}
\toprule
Module axis & $n$ & $r$ \\
\midrule
INT2/g16 & 12 & 0.412 \\
INT3/g16 & 15 & 0.35 \\
INT3/g32 & 15 & 0.415 \\
INT4/g16 & 15 & -0.043 \\
\bottomrule
\end{tabular}\quad
\begin{tabular}{lrr}
\toprule
INT2/g16 & Recon.\ err. & NDCG loss (\%) \\
\midrule
BGE-M3 & 0.3146 & 98.2 \\
E5-base-v2 & 0.3195 & 55.2 \\
EmbeddingGemma-300M & 0.3373 & 34.1 \\
Qwen3-Emb-0.6B & 0.3292 & 98.7 \\
SkillRet-0.6B (finetuned) & 0.3292 & 98.3 \\
\bottomrule
\end{tabular}\end{table}

Reconstruction error is the cheapest possible sensitivity proxy: quantize a tensor, measure
how far it moved, spend bits where it moved most. Table~\ref{tab:recon} splits its
predictive power by what is being varied. Throughout, the error of an arm is the
relative Frobenius error per quantized tensor, averaged with parameter-count weights
\emph{over the quantized tensors}: a whole-model figure for a uniform arm, and the module's
own error for a module arm --- the quantity an allocator compares when ranking modules. The
whole-model variant (the same number scaled by the fraction of parameters the arm touched)
is reported below wherever the two could disagree.

Along the uniform axis it works, and the relationship is monotone rather than merely linear:
excluding arms that lost more than half their quality, Pearson $r$ runs $0.80$ to $0.92$
across the five checkpoints and Spearman $\rho$ is $1.00$ for two of them and at least $0.89$
for the rest; including the collapsed arms raises every $r$ to $0.91$ or above. The lowest
per-model figure is E5-base-v2's $0.80$, with bootstrap interval $[0.48, 0.99]$. (Measured
without its prompt prefixes, as in the pre-fix run described in Section~\ref{sec:audit}, E5
had read $r=0.56$ with an uninformative interval.) For uniform bit-width screening, the proxy
is informative.

Along the module axis, pooling bit widths together is misleading. With the INT2 module arms of
Section~\ref{sec:int2mod} included, the pooled module correlation is $0.658$, but that rise
comes from range extension: bit width changes quantization severity at the same time module
identity changes, so the uniform axis leaks into the module figure. Holding the operating point
fixed removes the leak. Across the evaluated checkpoints the correlation is $-0.043$ at
INT4/g16, $0.350$ at INT3/g16, $0.415$ at INT3/g32 and $0.412$ at INT2/g16, against $0.88$ for
the uniform arms. The proxy does not become more informative as the allocation decision becomes
more consequential. Relaxing the \mbox{pre-specified} exclusion at INT2 --- adding back the three
module arms that lose more than half their quality --- moves that figure to $0.294$ rather than
improving it. The contrast survives the
obvious objection --- that excluding collapsed arms removes the non-linear region and
inflates the uniform figure --- because including those arms raises the uniform correlation
further, to $0.91$ or above in every model, while leaving the module arms where they are. The verdict does not depend on the denominator: scaling each module arm's error by the
fraction of parameters it touched, so that the axis reads whole-model error, moves the
within-bit correlations to $-0.272$, $-0.151$, $-0.150$ and $-0.061$ and the pooled module
figure to $0.413$, while the uniform axis stays at $0.88$ under either definition.

This matters because the two axes are not equally useful. Choosing a bit width is a
one-dimensional search a practitioner can run exhaustively. Choosing which tensors to protect
is the combinatorial problem that mixed-precision methods exist to solve, and it is the axis
on which the cheap proxy is weakest.

Across families the proxy fails outright where the decision matters. At INT2 reconstruction error is tightly
clustered between $0.315$ and $0.337$ across the five checkpoints, while retained NDCG ranges
from 1.3\% to 65.9\%.

\subsection{Task fine-tuning does not create the cliff}
\label{sec:paired}

Because a fine-tuned checkpoint appears in our set, the natural alternative explanation for
the divergence in Section~\ref{sec:recon} is training rather than architecture: perhaps
in-domain adaptation sharpens a representation in a way that makes it fragile. The paired
control tests this directly, since SkillRet-Embedding-0.6B differs from
Qwen3-Embedding-0.6B in training and nothing else.

One caution governs the comparison. SkillRet is in-domain for the fine-tuned checkpoint and
out-of-domain for the base, so retention measured there is not a like-for-like quantity; the
base model's SkillRet retention wanders above 100\% at one arm precisely because its
full-precision score on that corpus is low enough for noise to dominate. We therefore read
the verdict off SciFact and NFCorpus, which are out-of-domain for both.

\begin{table}[t]\centering\small
\caption{Paired control. Retained NDCG@10 (\% of full precision) for the base checkpoint and
for the same checkpoint fine-tuned on 127{,}190 in-domain pairs, averaged over SciFact and
NFCorpus --- both out-of-domain for both models, so the comparison is like-for-like. Task
fine-tuning moves quantization tolerance by at most 1.0 point in the two-corpus averages
shown here, and by at most 2.28 points on any individual model--corpus arm. In the averages
the movement is always toward greater robustness, but that is partly cancellation: at
INT3/g32 the fine-tune gains 2.28 points on SciFact and loses 2.08 on NFCorpus, which the
average reports as $+0.1$. Either way the base model already shows both the INT3 decline and
the INT2 collapse, so fine-tuning did not create them.}
\label{tab:paired}
\begin{tabular}{lrrrrrr}
\toprule
Checkpoint & W8 g16 & W4 g16 asym & W4 g128 asym & W3 g16 & W3 g32 & W2 g16 \\
\midrule
Qwen3-Emb-0.6B (base) & 99.7 & 99.9 & 96.2 & 91.5 & 89.6 & 1.6 \\
\quad + SkillRet fine-tune & 100.2 & 99.9 & 97.2 & 92.4 & 89.7 & 2.4 \\
\midrule
Difference & +0.5 & +0.0 & +1.0 & +0.9 & +0.1 & +0.7 \\
\bottomrule
\end{tabular}\end{table}

The two-corpus averages are small and non-negative, but the per-corpus effects are not
uniformly so: at INT3/g32 a $+2.28$-point change on SciFact and a $-2.08$-point change on
NFCorpus cancel to $+0.1$ in the average. More importantly, the base checkpoint already exhibits both
phenomena the fine-tuned one does: an INT3 decline to 91.5\% and an INT2 collapse to 1.6\%.
Task adaptation is not what puts the cliff there.

\subsection{NDCG@10 does not see what quantization moves}
\label{sec:rank}

\begin{table}[t]\centering\small
\caption{Retrieval quality against ranking stability, averaged over the corpora measured for
each model. Retained NDCG@10 is the percentage of the full-precision score; top-10 overlap is
the fraction of the full-precision top-10 still present; the flip rate is the fraction of
queries whose rank-10 document left the top-10 entirely. At INT4/g16 the metric barely moves
--- in two cells it reports a gain --- while 13\% to 22\% of the top-10 has been replaced
(11.6\% to 25.2\% across individual model--corpus cells) and the boundary document moves for
roughly half the queries. E5-base-v2 is absent from this table because the rank-preservation
arm was never measured for it; its eviction composition, which was
measured, is in Table~\ref{tab:turnover} --- where what is displaced is counted against the
relevance judgements rather than inferred from the metric here.}
\label{tab:rank}
\begin{tabular}{lrrr}
\toprule
Arm & Retained NDCG@10 (\%) & Top-10 overlap & Rank-10 flip rate \\
\midrule
\multicolumn{4}{l}{\emph{Qwen3-Emb-0.6B}} \\
\quad W8 g16 & 99.8 & 0.967 & 0.225 \\
\quad W4 g16 asym & 100.4 & 0.796 & 0.555 \\
\quad W3 g16 & 89.6 & 0.577 & 0.688 \\
\quad W2 g16 & 1.3 & 0.005 & 0.995 \\
\multicolumn{4}{l}{\emph{SkillRet-0.6B (finetuned)}} \\
\quad W8 g16 & 100.1 & 0.973 & 0.208 \\
\quad W4 g16 asym & 99.8 & 0.805 & 0.534 \\
\quad W3 g16 & 92.2 & 0.589 & 0.699 \\
\quad W2 g16 & 1.7 & 0.004 & 0.999 \\
\multicolumn{4}{l}{\emph{EmbeddingGemma-300M}} \\
\quad W8 g16 & 100.0 & 0.990 & 0.088 \\
\quad W4 g16 asym & 99.9 & 0.872 & 0.484 \\
\quad W3 g16 & 97.6 & 0.745 & 0.579 \\
\quad W2 g16 & 65.7 & 0.329 & 0.842 \\
\multicolumn{4}{l}{\emph{BGE-M3}} \\
\quad W8 g16 & 100.1 & 0.981 & 0.158 \\
\quad W4 g16 asym & 98.7 & 0.778 & 0.551 \\
\quad W3 g16 & 94.0 & 0.567 & 0.703 \\
\quad W2 g16 & 1.8 & 0.004 & 0.996 \\
\bottomrule
\end{tabular}\end{table}

\begin{figure}[t]\centering
\includegraphics[width=\linewidth]{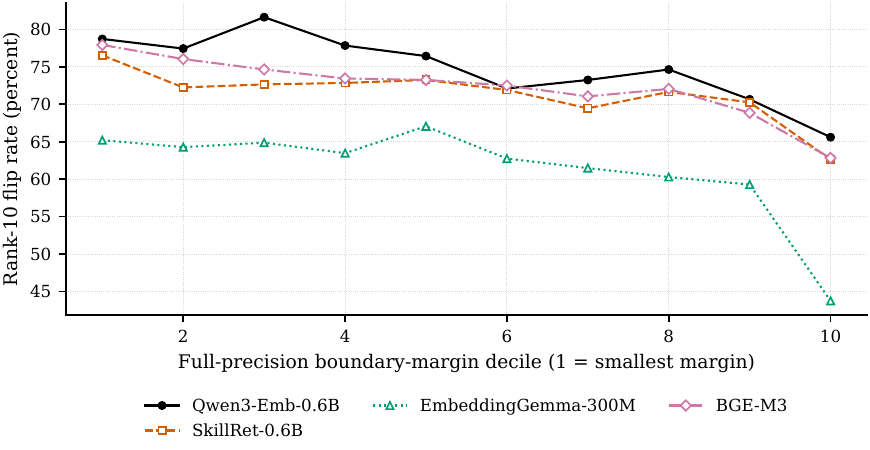}
\caption{Fraction of queries whose rank-10 document leaves the top-10 after INT3/g16
quantization, binned by the full-precision model's boundary margin (the score gap between its
10th and 11th documents), weighted across corpora. Smaller margins do flip more often, but
the relationship is shallow: the spread within deciles one to nine is about as large as the
gap between the first decile and the tenth. E5-base-v2 is omitted because the
rank-preservation arm this analysis needs was not measured for that checkpoint, the same
reason it is absent from Table~\ref{tab:rank}.}
\label{fig:flip}\end{figure}

Every number to this point is NDCG@10, the benchmark's headline metric. It is also the metric
that decides whether a practitioner calls a quantization ``free''. Table~\ref{tab:rank} adds
two measurements NDCG does not contain: how much of the full-precision top-10 survives, and
how often the document sitting exactly at rank 10 is pushed out of the list entirely.

At INT4/g16 the metric barely moves: no checkpoint loses as much as a single NDCG point in
its three-corpus average (Table~\ref{tab:modules}), and retention runs from 98.7\% to 100.4\%, the latter above
full precision. This is where we stop calling INT4/g16 ``free''. It is NDCG-neutral at this
operating point; it is not ranking-invariant. Underneath that, between 13\% and 22\% of the top-10 has been
replaced, and the rank-10 document leaves the list for roughly half of all queries. The
ranking a user would see is visibly different and the score is not.

The natural worry is that NDCG is simply insensitive, and the natural rebuttal --- that a
drop would have shown up if relevant documents had moved --- is an inference, not a
measurement. It is also not airtight: a query with several gold documents can lose one and
gain another at no cost to NDCG, and per-query gains and losses can cancel in the mean. Since
we hold the relevance judgements, we count instead.

\begin{table}[t]\centering\small
\setlength{\tabcolsep}{4pt}
\caption{What quantization actually displaces, counted against the relevance judgements
rather than inferred from NDCG. ``Evicted'' counts documents leaving the full-precision
top-10, summed over each model's corpora; the next column is the relevant share of those.
\textbf{Read the last column, not the eviction share.} On a hard corpus everything churns,
so the relevant share rises for healthy and broken models alike --- at INT2 on NFCorpus it is
22\% for EmbeddingGemma, which keeps two thirds of its NDCG, and 23--26\% for BGE-M3 and
Qwen, which keep none. What separates them is how much of the gold set survives: a
relevant-only overlap of 0.500 against 0.000. At INT4/g16 that overlap is 0.944 to 0.970 in the per-model averages shown here, and 0.899
to 1.000 across individual model--corpus cells
and recall moves by less than a point, which is the sense in which INT4 leaves the gold set
alone while replacing a sixth of the list.}
\label{tab:turnover}
\begin{tabular}{lrrrrr}
\toprule
Arm & Ret.\ NDCG (\%) & Evicted & \ldots relevant & $\Delta$Recall@10 & Rel.-only overlap \\
\midrule
\multicolumn{5}{l}{\emph{Qwen3-Emb-0.6B}} \\
\quad W8 g16 & 99.8 & 2034 & 2.0\% & +0.10 & 0.991 \\
\quad W4 g16 asym & 100.4 & 11964 & 2.1\% & +0.58 & 0.952 \\
\quad W3 g16 & 89.6 & 25288 & 4.0\% & -4.18 & 0.843 \\
\quad W2 g16 & 1.3 & 49933 & 10.6\% & -54.37 & 0.006 \\
\multicolumn{5}{l}{\emph{SkillRet-0.6B (finetuned)}} \\
\quad W8 g16 & 100.1 & 1387 & 2.5\% & -0.03 & 0.993 \\
\quad W4 g16 asym & 99.8 & 10204 & 2.4\% & -0.27 & 0.952 \\
\quad W3 g16 & 92.2 & 22714 & 4.0\% & -3.83 & 0.860 \\
\quad W2 g16 & 1.7 & 49974 & 14.0\% & -60.95 & 0.006 \\
\multicolumn{5}{l}{\emph{EmbeddingGemma-300M}} \\
\quad W8 g16 & 100.0 & 520 & 1.9\% & +0.01 & 0.999 \\
\quad W4 g16 asym & 99.9 & 7243 & 2.4\% & +0.08 & 0.970 \\
\quad W3 g16 & 97.6 & 14800 & 3.0\% & -1.29 & 0.926 \\
\quad W2 g16 & 65.7 & 36865 & 6.3\% & -17.19 & 0.620 \\
\multicolumn{5}{l}{\emph{BGE-M3}} \\
\quad W8 g16 & 100.1 & 976 & 2.3\% & +0.17 & 0.994 \\
\quad W4 g16 asym & 98.7 & 12368 & 1.9\% & -0.66 & 0.944 \\
\quad W3 g16 & 94.0 & 24135 & 3.1\% & -2.73 & 0.853 \\
\quad W2 g16 & 1.8 & 49940 & 10.2\% & -51.54 & 0.001 \\
\multicolumn{5}{l}{\emph{E5-base-v2}} \\
\quad W8 g16 & 100.0 & 687 & 1.7\% & -0.03 & 0.998 \\
\quad W4 g16 asym & 99.0 & 8902 & 2.6\% & -0.90 & 0.952 \\
\quad W3 g16 & 98.5 & 18921 & 2.2\% & -0.76 & 0.910 \\
\quad W2 g16 & 41.9 & 41116 & 6.4\% & -24.79 & 0.409 \\
\bottomrule
\end{tabular}\end{table}

Table~\ref{tab:turnover} resolves it. Its shares pool every eviction, which weights corpora
by how much they churn; averaging the three corpora equally instead moves the INT4/g16 relevant
share from 2.2\% to 5.2\% and the INT2 share from 9.8\% to 13.4\%. The gap between the two
aggregations is itself the corpus variance discussed below, and we quote both. At INT4/g16 the documents leaving the top-10 are almost
entirely non-relevant, recall@10 moves by less than a point on average across corpora, and the overlap computed over relevant
documents alone stays near one --- the gold set is essentially untouched while the list around
it churns. The concentration is not uniform across corpora, and the spread is
worth stating: for EmbeddingGemma at INT4/g16 the relevant share of evictions is 0\% on SciFact
(0 of 347), 1.7\% on SkillRet (113 of 6{,}506) and 15.1\% on NFCorpus (59 of 390). NFCorpus
is the hardest of the three for every model we tested, and it is where the tail and the gold
set are least separable. Reporting only the cleanest corpus would overstate the effect.

The measurement also discriminates, and in doing so bounds its own scope. It also shows which
column to trust. On a hard corpus the relevant share of evictions stops separating healthy
models from broken ones --- at INT2 on NFCorpus it is 22\% for EmbeddingGemma, which retains
two thirds of its NDCG, against 23\% and 26\% for BGE-M3 and Qwen, which retain none. The
relevant-only overlap does separate them, at 0.500 against 0.000 and 0.011. When everything
churns, what matters is not how many gold documents were touched but how many stayed. At INT2, where NDCG
does collapse, evictions include relevant documents and recall drops steeply. The two
outcomes at that bit width are not the same failure. On SciFact, EmbeddingGemma --- which
retains 69\% of its NDCG --- keeps a relevant-only overlap of 0.764, while Qwen, which
retains none, keeps \emph{nothing}, at 0.000. Across all three corpora EmbeddingGemma's INT2
relevant-only overlap stays between 0.500 and 0.764 where Qwen's is essentially zero.
Surviving INT2 turns out to mean holding on to the gold set specifically.

So the earlier inference survives contact with the data, but with a limit we had not
anticipated and would not have found by reasoning: damage concentrates in the non-relevant
tail only while the model is still working. Past its own cliff what breaks is not the
concentration itself --- non-relevant evictions still dominate --- but the preservation of the
gold set.

This cuts against the motivation for boundary-preservation objectives more sharply than
anything else we measured. The target here is an objective that tries to hold the teacher's
entire ranking, including reorderings the metric does not score; it is not an argument
against relevance-aware ranking losses of the kind JPQ~\cite{jpq} and RepCONC~\cite{repconc}
train against, which optimise for relevance rather than for teacher-order stability. A loss that penalises boundary reordering would spend its capacity
defending precisely the positions whose movement leaves the metric unchanged.
Figure~\ref{fig:flip} adds a second constraint on such a design: the boundary margin does
predict flips, with a rank correlation around $-0.6$ to $-0.8$, but the curve is shallow
enough that margin-proportional weighting would be close to uniform weighting outside the
top decile.

\FloatBarrier
\section{Controls, Reproducibility, and Cliff Probes}
\label{sec:repro}

Every gate below exists because the corresponding failure occurred in our own measurements
and produced a number we briefly believed.

\paragraph{Prompt hash.} Training a checkpoint on bare queries and evaluating it with a
20-token instruction prefix cost 8.84 points of NDCG@10 on one checkpoint and 4.23 on
another. Worse, it did not merely lower the scores: quantization noise partially undid the
mismatch, so the quantized arm scored \emph{above} the full-precision arm, and the apparent
gain grew with training epochs, which made it look like a mechanism. We retracted that
result. Our evaluation bundle stores a \texttt{query\_prefix.json} for every checkpoint recording the resolved
prefix, and an evaluator that cannot find one refuses to run. Note that \emph{which} prefix
is used barely matters after fine-tuning; only consistency does.

\paragraph{Byte ledger.} A size claim is arithmetic until a file of that size exists. Every
reported size is the byte count of an actual on-disk artifact: serialized weights for the
full-precision rows, and for the low-bit rows the packed quantized file rather than an
arithmetic estimate. The arithmetic prediction runs 0.04 to 0.45\% above the packed file across the seven quantized artifacts, so the two are close but not interchangeable; the measured byte count is what we report, and \texttt{artifacts.csv} in the measurement repository carries the file name, byte count and SHA-256 behind every size in this paper.

\paragraph{Holdout isolation.} Splitting by a query's first gold document leaves every
``held-out'' document reachable through that query's other golds. In our case this left
100\% of the nominally isolated documents in training and inverted a model-selection
decision. Splits are taken over all golds.

\paragraph{Fake quantization is not a kernel.} All quality numbers here are fake-quantized
with FP32 arithmetic. We make no latency claim from them.

\subsection{Vocabulary size and embedding-table fragility do not explain the INT2 cliff}
\label{sec:int2mod}

The module arms in Section~\ref{sec:modules} run at INT4 and INT3. That is every regime
except the one where models actually break, which makes the cliff the one place our
module-level claims were untested. An obvious candidate is vocabulary size and the
vocabulary-dependent embedding table it produces: the five checkpoints carry vocabularies from
30{,}522 to 262{,}144, vocabulary size sets the number of embedding rows, and the table is
21\% to 66\% of their parameters.

Two measurements argue against that explanation. First, vocabulary size does not order the outcome. Ranked by
vocabulary the survivors sit at both extremes --- E5-base-v2 at 30{,}522 keeps 44.8\% and
EmbeddingGemma at 262{,}144 keeps 65.9\% --- while BGE-M3 at 250{,}002 keeps 1.8\%. BGE-M3 and
EmbeddingGemma have similarly large vocabularies, 250{,}002 and 262{,}144, yet their INT2
retention differs by 64.1 points.

Second, and directly: we quantized \emph{only} the embedding table to INT2
(Table~\ref{tab:int2mod}). If vocabulary-dependent embedding-table fragility were the source of
the cliff, quantizing that table alone should reproduce substantial degradation. It does not. Embedding-only INT2 leaves 97.6\% to 100.5\% of retrieval quality
in all five checkpoints, including the three that keep under 2\% when everything is
quantized; BGE-M3 reads 100.5\% on it. The embedding table is not the dominant isolated
source of damage, and what replaces it is family-dependent --- not merely which module, but how
much of the failure any single module accounts for. BGE-M3's
attention alone \emph{retains} only 2.7\%, close to the 1.8\% it retains under full-model
INT2, and that holds in each corpus separately (0.20 against 0.00, 5.57 against 4.85, 2.24
against 0.41), so almost all of its collapse is reproduced by a single module. The
feed-forward block is instead the dominant isolated source of damage in the Qwen lineage, at
11.3\% and 9.9\% retained --- though neither reproduces the 1.3\% and 1.7\% of the joint arm,
which is the non-additivity we report throughout.

Computing the interaction residual of Section~\ref{sec:modules} at this operating point is
only partly meaningful. For the two checkpoints that remain above the floor the residual is
negative --- $-8.3$ points for EmbeddingGemma and $-8.8$ for E5-base-v2 --- so joint
degradation substantially exceeds the sum of the isolated effects. These residuals are computed
from absolute NDCG-point changes under the same definition as Table~\ref{tab:modules}, not
from the retention percentages in Table~\ref{tab:int2mod}. For the three that collapse the sum of
the single-module losses already exceeds the retrieval quality available --- there is no
further quality to lose --- so the residual comes out positive ($+7.3$ to $+10.0$) as a floor
artifact, and we do not read its sign.

This removes one confound from the list in Section~\ref{sec:limitations} rather than
explaining the cliff. What the intervention changes is the precision of the embedding table,
not the tokenizer: vocabulary size and embedding-table fragility cannot by themselves explain the
observed cliff, while tokenization behaviour itself --- how finely a tokenizer segments our
corpora, how much of its vocabulary they exercise, how much sequence length inflates ---
remains untested.

\begin{table}[t]\centering\small
\caption{Retained NDCG@10 (\% of full precision, averaged over the three corpora) when only
the named module is quantized to INT2/g16, against quantizing everything. Vocabulary size
directly determines the number of embedding rows, so this arm tests whether fragility of the
vocabulary-dependent embedding table explains the cliff; it does not test tokenization
behaviour itself. It does not explain it: embedding-only INT2 leaves 97.6\% to 100.5\% in the
per-model averages (94.7\% to 101.3\% across cells), including in the three checkpoints that
retain under 2\% when everything is quantized, and BGE-M3 reads 100.5\%. The decomposition of
the failure is itself family-dependent: BGE-M3 is close to a single-module bottleneck, since
quantizing its attention alone leaves almost what quantizing everything leaves; the Qwen
lineage is feed-forward-dominant with a large residual on top; and for EmbeddingGemma and
E5-base-v2 no isolated module comes near the joint loss at all. The All column is the same arm as in
Table~\ref{tab:uniform} and is read from the same source; the run reported here
re-measured it and agreed to within 0.05 points of retained quality.}
\label{tab:int2mod}
\begin{tabular}{lrrrr}
\toprule
Model & All & Embed.\ only & Attn.\ only & FFN only \\
\midrule
Qwen3-Emb-0.6B & 1.3 & 99.1 & 75.0 & 11.3 \\
SkillRet-0.6B (finetuned) & 1.7 & 99.7 & 74.7 & 9.9 \\
EmbeddingGemma-300M & 65.9 & 97.6 & 91.7 & 90.4 \\
BGE-M3 & 1.8 & 100.5 & 2.7 & 84.9 \\
E5-base-v2 & 44.8 & 97.6 & 81.0 & 84.3 \\
\bottomrule
\end{tabular}\end{table}

\subsection{What sets the INT2 cliff is not pooling}
\label{sec:pooling}

INT2 retention spans the full range rather than splitting in two. EmbeddingGemma keeps
65.9\% of its full-precision NDCG, E5 keeps 44.8\%, and the remaining three keep 1.3\% to
1.8\%. E5 sits between the extremes on the ranking measurements as well --- at INT2 on SciFact
its relevant-only top-10 overlap is 0.530 against 0.764 for EmbeddingGemma and 0.000 for both
collapses --- so ``survives'' and ``collapses'' are convenient labels for the ends of a
continuum, not two classes. Only the bottom three are at the floor.

With that caveat, the ordering lines up with pooling: the two models that retain anything
mean-pool, and the three that do not read a single position, CLS or last token.
The mechanism is easy to state --- mean pooling averages per-token quantization noise over
many positions while single-position pooling takes it undiluted --- and with five of five
checkpoints separating cleanly it is tempting to report.

We tested it by manipulation instead. Forcing BGE-M3 and Qwen3-Embedding to mean pooling
moves INT2 retention from 2.4\% to 4.1\% and from 1.6\% to 9.5\%, far short of the
surviving checkpoints. Removing mean pooling from EmbeddingGemma leaves it at 30.5\%, still far
above any of the collapses. Mean pooling is therefore not sufficient for INT2 robustness. Whether it is necessary
remains unresolved, because the reverse intervention destroys the representation it acts on
(below).

These pooling interventions are reported as averages over SciFact and NFCorpus, matching the
out-of-domain two-corpus protocol used for the paired control; that is why the baselines here
differ from the three-corpus retention figures quoted above.

The manipulation is not clean in both directions. Imposing a pooling a model was not trained
for costs it most of its full-precision quality --- EmbeddingGemma falls from 58.5 to 4.2 ---
so the 30.5\% figure sits on top of an already broken model and should be read weakly. The
direction that survives is the one that matters for the claim: mean pooling, given to a model
that collapses, does not rescue it.

We therefore do not know what separates the two groups. We prefer saying so to naming a
factor that the sufficiency test refutes and the necessity test cannot decide.

\FloatBarrier
\section{Discussion: distil small, or squeeze large?}
\label{sec:discussion}

Under the group-wise affine weight-only PTQ regime we test, a 0.6B embedder of the
Qwen lineage collapses below three bits. It does not say
what to do instead, and the obvious alternative is to train a smaller model and quantize it
gently. We have two such checkpoints --- 22M and 109M students distilled from the
SkillRet-finetuned 0.6B --- and evaluating them on the same three corpora with the same gate
gives a sharper answer than either the PTQ result or the distillation result alone.

\begin{table}[t]\centering\small
\caption{Distilled students against general-purpose embedders, same gate and same corpora.
SkillRet is the domain the students were distilled on; SciFact and NFCorpus are not. Inside
the domain the 109M student at INT3 holds 78.04 in 68.4\,MB. Outside it the same checkpoint
scores 7.79 on NFCorpus, against 31 to 39 for models that were never specialised. The distilled student strictly dominates the extreme-PTQ arm on the
size--quality frontier --- 68.4\,MB at 78.04 against 297.9\,MB at 64.46 --- but only within
the task it was trained for. Three precisions are distinct here and should not be conflated: every score in
this paper is computed in FP32 arithmetic; the teacher's reference storage artifact is its
1191.6\,MB BF16 checkpoint as distributed; and student sizes are measured on-disk artifacts:
serialized FP16 for the FP16 rows, packed quantized files for the INT rows. The lower block holds the 0.6B teacher, which is specialised for SkillRet, and
four general-purpose references, which are not; size is left blank for the four, which are
included for their scores rather than as size comparisons. $^{\dagger}$BF16 storage
artifact as distributed; evaluation uses FP32 arithmetic like every other row.}
\label{tab:students}
\begin{tabular}{lrrrr}
\toprule
Model & Size (MB) & SkillRet & SciFact & NFCorpus \\
\midrule
22M student, FP16 & 45.4 & 75.26 & 55.58 & 16.86 \\
22M student, INT4/g16 & 17.0 & 75.14 & 56.40 & 17.29 \\
22M student, INT3/g16 & 14.2 & 73.82 & 56.78 & 16.90 \\
109M student, FP16 & 219.0 & 79.18 & 50.17 & 7.54 \\
109M student, INT4/g16 & 82.0 & 79.21 & 53.07 & 7.31 \\
109M student, INT3/g16 & 68.4 & 78.04 & 50.73 & 7.79 \\
\midrule
SkillRet-0.6B teacher, INT3/g32 & 297.9 & 64.46 & 64.94 & 29.10 \\
SkillRet-0.6B teacher$^{\dagger}$ & 1191.6 & 78.48 & 69.43 & 33.86 \\
Qwen3-Emb-0.6B, FP32 & -- & 57.48 & 70.32 & 35.40 \\
EmbeddingGemma-300M, FP32 & -- & 60.44 & 78.32 & 38.75 \\
BGE-M3, FP32 & -- & 56.25 & 64.37 & 31.57 \\
E5-base-v2, FP32 & -- & 51.72 & 71.05 & 35.52 \\
\bottomrule
\end{tabular}\end{table}

Inside the distillation domain the case is strong. The 109M student at INT3/g16 occupies
68.4\,MB and scores 78.04 on SkillRet, within 0.44 points of the 0.6B teacher's
full-precision 78.48, at roughly one seventeenth the size of the teacher's 1191.6\,MB BF16
checkpoint as distributed, which is the only full-precision reference artifact we use for
storage comparisons; no FP32 checkpoint file was written, so we state no ratio against FP32
weights, a size claim being arithmetic until a file of that size exists; in full precision it scores 79.18,
slightly \emph{above} the teacher. The student is therefore not a compressed copy of the
teacher but a different in-domain Pareto point, bought with the generality lost below. None of our PTQ arms on the 0.6B reaches that
point: its own INT3/g32 checkpoint is larger and scores 64.46.

Outside that domain the same checkpoints fall apart. The 109M student scores 7.79 on NFCorpus
where every general-purpose model we measured scores between 31 and 39, and 50.73 on SciFact
against 64 to 78. The 22M student is less extreme but shows the same shape --- and the degradation is not
monotonic in student size. The 22M student generalises \emph{better} than the 109M on both
external corpora (16.9 against 7.5 on NFCorpus, 55.6 against 50.2 on SciFact) while being
weaker in-domain. With two students we do not attribute this to capacity, but it cautions
against treating distillation size as a simple quality knob. These students
were distilled on SkillRet pairs alone, so the collapse may reflect the training distribution
rather than distillation as such --- but as measured, they are not general embedders.

The practical conclusion is therefore narrower than ``distil small, then quantize''. For a
\emph{fixed, known} retrieval task, training a small student and quantizing it moderately
dominates the extreme-PTQ point we measured on the size--quality Pareto frontier: 68.4\,MB at
78.04 against 297.9\,MB at 64.46. For a general-purpose embedder the
comparison does not hold, and the size--quality argument is only legitimate within the task
the student was trained for.

\subsection{Audit of the measurement code}
\label{sec:audit}

Before releasing this paper we audited the measurement code against the text, applying the module masks to the actual tensor names of every checkpoint, probing
the quantizer on synthetic tensors, and recomputing every quoted statistic from the raw
per-query scores. The masks, the quantizer, the pooling and the statistics reproduce the
tables; the following deviations between an earlier draft and the implementation were found,
and their consequences are recorded here. (i) The reconstruction error of a module arm was
averaged over the quantized tensors only, not over the whole model as an earlier draft
implied; Section~\ref{sec:recon} states the definition and reports both variants, and a
sentence comparing a whole-model figure with a module figure was withdrawn. (ii) The
evaluator applied a model's prompt to queries only. Qwen3-Embedding and its fine-tune have an
empty document prompt and BGE-M3 uses none, so they are unaffected; E5-base-v2 ships no
sentence-transformers prompt configuration and was therefore evaluated without its
\texttt{query:}/\texttt{passage:} prefixes, and EmbeddingGemma without its document prefix.
Both were re-measured under their own prompt contracts with the corrected evaluator, and every
E5-base-v2 and EmbeddingGemma number in this paper comes from that re-measurement.
Full-precision NDCG@10 moved by at most 1.4 points relative to the pre-fix run. The largest
downstream changes were E5's INT2 retention (41.9\% to 44.8\%); E5's uniform-axis
correlation, which the pre-fix run had put at $r=0.56$ with an uninformative interval and
which reads $0.80$ under the prompt contract (Section~\ref{sec:recon}); and the INT3/g16
module ordering of the two re-measured checkpoints, where EmbeddingGemma's attention (0.71
points) edges the feed-forward block (0.56) that the pre-fix run had named as costliest, and
E5, tied in the pre-fix run, is attention-dominant. Eight rather than six of the forty-five INT4 module cells survive the
Benjamini--Hochberg correction, and four rather than three of the ten interval-excluding cells
are positive. Every headline verdict --- the INT4 headroom bound of 1.01 points, the
family-dependent INT3 ordering, the sign instability of the interaction residual, the
within-bit correlations below 0.42 and the INT2 divergence --- is the same under either run. Three measurements
were not repeated and remain from the pre-fix evaluator: the top-10 turnover of
Table~\ref{tab:turnover} for E5 and EmbeddingGemma, EmbeddingGemma's rank-flip curve, and the
pooling intervention of Section~\ref{sec:pooling}; each compares arms of one model under one
constant prompt convention, so the comparison stands but the absolute values are pre-fix.
(iii) The quantizer walks the transformer only; EmbeddingGemma's two sentence-transformers
dense output layers (4.7M parameters) stayed at full precision in every arm. Quantizing them
together with everything else changes EmbeddingGemma's INT3/g16 retention by 0.1 points and
its INT2 retention by 1.1 points (64.8\% against 65.9\%), so the unquantized output path does
not explain why it survives INT2. (iv) The uniform arm
quantizes position embeddings that no module arm touches (1.5\% of BGE-M3's and 0.4\% of
E5's parameters), so the interaction residual for those two checkpoints includes that
leftover; the residual is exactly attributable for the other three. (v) NFCorpus relevance
judgements were binarised. (vi) An earlier draft's correlation table mixed three
generations of its underlying rows: the per-model and robustness figures predate the INT2
module arms, and the pooled module figure of $0.658$ predates the backfilled INT3/g32
attention arm and so omitted five rows. Every block of Table~\ref{tab:recon} is computed
on the same ninety-two rows; the pooled module figure is $0.658$ on that set as well, by
coincidence rather than by construction.

\section{Limitations}
\label{sec:limitations}

The evaluation covers three retrieval corpora, not full MTEB, and no multilingual retrieval
benchmark; BGE-M3's multilingual capability is therefore not exercised. We measure
weight-only PTQ and do not quantize activations. We do not implement or evaluate a
mixed-precision allocator: our result is that at INT4/g16 there is no large or consistently
ordered sensitivity differential for one to exploit, which is an argument about headroom rather than a
demonstration that a specific allocator fails. We report no integer-kernel latency, and low-bit storage reduction does not imply
proportional CPU speedup --- closing that gap is a separate systems problem addressed by
specialised kernels such as T-MAC~\cite{tmac}. Our INT2 embedding-table intervention shows that
vocabulary-dependent table fragility cannot by itself explain the cliff; it says nothing about
tokenization behaviour: corpus-specific segmentation,
how much of a vocabulary an evaluation set actually exercises, and sequence-length inflation all
remain unmeasured. Finally,
the cross-family divergence rests on four families, not five checkpoints --- the
fifth is a paired control within one of them --- and four families is a small basis for a
claim about embedders in general. It establishes that a policy measured on one family failed
to describe another, not which architectural property causes the difference, and not that no
further family behaves like a fourth case we did not test.

\section{Conclusion}

We began intending to build a retrieval-aware mixed-precision PTQ method for text embedders,
and measured its premises before building it. None of the individual phenomena we report is
unprecedented: sensitivity-guided allocation dates to HAWQ~\cite{hawq,hawqv2}, non-separable
quantization loss at aggressive bit widths to LAPQ~\cite{lapq}, and cross-layer dependency to
BRECQ~\cite{brecq}. What we add is that for retrieval embedders, judged by retrieval order
rather than reconstruction or perplexity, the usefulness of the standard sensitivity proxy
and the ordering of module sensitivities both change with the operating bit width, and
what changes is not simply the proxy's accuracy but where it is useful: reconstruction error
transfers reasonably well for uniform severity sweeps and is substantially less reliable for
module selection, while module-sensitivity ordering itself does not transfer reliably across
families. The measurements neither cleanly confirmed nor
cleanly refuted them; they relocated the problem.

A cheap reconstruction proxy is useful for screening uniform bit widths but substantially
less reliable for choosing which tensors to protect, so if a retrieval-aware sensitivity signal is worth
building, the module axis is where it has to prove itself --- not the uniform sweep where
such methods are usually demonstrated. The same measurements narrow what any such method can
promise. At INT4/g16 the module-level sensitivity that exists is too small to allocate against. At INT3
there is, and its ordering inverts between families. Per-module costs fail to predict joint cost
additively; both the magnitude and, at INT3/g16, even the sign of the interaction vary across
model families. And at INT2 five checkpoints whose reconstruction error
spans $0.315$ to $0.337$ retain between 1.3\% and 65.9\% of their retrieval quality.

The recurring finding is not that compression is hard but that its policy does not transfer
reliably.
Every heuristic we tested failed to transfer as stated. The embedding table never emerged as the
dominant isolated protection priority in any of the four families, despite being the largest
tensor in several of them --- the rule is not wrong so much as pointed the wrong way. Protecting the feed-forward
block was right in some families and wrong in others, reconstruction error was trustworthy on
one search axis and not the other, and INT3 robustness varied sharply by family.

\bibliographystyle{plain}
\bibliography{refs}

\end{document}